\documentclass[11pt,letterpaper]{article}
\usepackage[margin=1in]{geometry}
\usepackage[T1]{fontenc}
\usepackage{lmodern}
\usepackage{microtype}
\usepackage{amsmath,amssymb,amsthm,mathtools}
\usepackage{enumitem}
\usepackage{booktabs}
\usepackage{tikz}
\usepackage{xcolor}
\usepackage[colorlinks=true,linkcolor=blue!55!black,citecolor=blue!55!black,urlcolor=blue!55!black]{hyperref}
\usepackage[nameinlink,capitalise,noabbrev]{cleveref}
\hypersetup{pdftitle={A (12/11 + epsilon)-Approximation for Nonpreemptive Throughput Maximization},pdfauthor={Mikhail Cherniavskii, Mikhail Koshelev, Andrey Kupavskii, Stanislav Moiseev, Fedor Ozhegov, Alexander Shekhovtsov},pdfkeywords={scheduling, throughput maximization, approximation algorithms}}
\allowdisplaybreaks[1]
\setlist{itemsep=3pt,topsep=5pt}
\newtheorem{theorem}{Theorem}[section]
\newtheorem{lemma}[theorem]{Lemma}
\newtheorem{proposition}[theorem]{Proposition}

\theoremstyle{definition}

\theoremstyle{remark}

\title{A $(12/11+\varepsilon)$-Approximation for\\Nonpreemptive Throughput Maximization}
\author{Mikhail Cherniavskii \qquad Mikhail Koshelev \qquad
Andrey Kupavskii \\ Stanislav Moiseev \qquad
Fedor Ozhegov \qquad Alexander Shekhovtsov}
\date{}
\begin{document}
\maketitle
\begin{abstract}
We study unweighted nonpreemptive throughput maximization on an arbitrary
number of identical machines:
each job has a processing time and a release--deadline window, and the goal is
to schedule as many jobs as possible. For every fixed $\varepsilon>0$, we give
a randomized $(12/11+\varepsilon)$-approximation. This improves the recent $(4/3+\varepsilon)$-approximation of Alexander Armbruster,
Fabrizio Grandoni, Antoine Tinguely, and Andreas Wiese~\cite{AGTW26}
and the $(17/13+\varepsilon)$-approximation of Tong and Xu~\cite{TX26}.
For binary-encoded integral inputs, our algorithm runs in time
$n^{O_\varepsilon(1)}\operatorname{poly}(N)$, where $N$ is the input length. 
\end{abstract}
\clearpage

\section{Introduction}\label{sec:introduction}
A job with a release time and a deadline may have many possible execution
intervals, but it contributes only once to a schedule. This distinction is
central to throughput maximization. A fractional solution may place a job
in several parts of the time horizon. Sampling these placements independently
can produce repeated copies of that job while leaving another job absent.
The resulting execution slots still have value: a spare copy can sometimes
be replaced by a missing job. The difficulty is to recover this value while
respecting both slot capacities and the requirement that each job count once.

We consider the offline, unweighted problem on $M$ identical machines,
where $M$ is part of the input. The input also contains a set $J$ of $n$ jobs. Job $j$ has an integral processing time $p_j>0$, release
time $r_j$, and deadline $d_j$. A feasible schedule selects jobs and assigns each selected job to one
machine and one execution interval of length $p_j$ contained in
$[r_j,d_j)$; intervals assigned to the same machine are pairwise disjoint. The objective is the number
of selected jobs. Write $\mathrm{OPT}$ for the optimum and $N$ for the total
binary encoding length. Processing times and window lengths are unrestricted,
and no resource augmentation is allowed.

The block-configuration approach of
Armbruster, Grandoni, Tinguely, and Wiese~\cite{AGTW26}, building on
Chuzhoy, Ostrovsky, and Rabani~\cite{COR06}, supplies our starting relaxation.
After a small loss, the analysis may restrict attention to a schedule whose
executions lie inside blocks, each containing a bounded number of jobs.
Sampling a feasible configuration per block produces disjoint execution
slots. A maximum matching then assigns compatible jobs to those slots,
allowing a slot to serve a job other than the one that generated it.
Tong and Xu~\cite{TX26} strengthen this approach through crossing exchanges
and coordinated boundary choices. We improve the polynomial-time
approximation guarantee for an arbitrary number of identical machines.

\subsection{Result and comparison}
An approximation factor $\rho\ge 1$ means expected cardinality at least
$\mathrm{OPT}/\rho$.
\begin{theorem}\label{thm:main}
For every fixed $\varepsilon>0$, there is a randomized algorithm for
unweighted nonpreemptive throughput maximization on an arbitrary number
of identical machines with binary-encoded integral input data that runs in time
$n^{O_\varepsilon(1)}\operatorname{poly}(N)$ and always returns a feasible
schedule whose expected cardinality is at least
$\mathrm{OPT}/(12/11+\varepsilon)$.
\end{theorem}
Feasibility holds for every outcome of the algorithm. The expectation concerns
only the number of scheduled jobs. In particular, the running time is
polynomial in the binary input length for every fixed accuracy, uniformly
over the number of machines. We develop the proof for one machine in
\cref{sec:relaxation,sec:coordination,sec:matching,sec:main-proof}.
Appendix~\ref{sec:multiple-machines} extends the construction to every
bounded number of machines and combines it with the large-machine
algorithm of Im, Li, and Moseley~\cite{ILM20}.

\begin{table}[!ht]
\centering
\small
\setlength{\tabcolsep}{5pt}
\begin{tabular}{@{}lcc@{}}
\toprule
Result & Identical machines & Approximation factor\\
\midrule
Armbruster et al.~\cite{AGTW26} & Arbitrary & $4/3+\varepsilon$\\
Tong--Xu~\cite{TX26} & Arbitrary & $17/13+\varepsilon$\\
\cref{thm:main} & Arbitrary & $12/11+\varepsilon$\\
\bottomrule
\end{tabular}
\caption{Recent polynomial-time guarantees for unweighted nonpreemptive
throughput maximization with arbitrary processing times. All listed factors
are in expectation, without resource augmentation. For fixed
$\varepsilon>0$, the running times are
$n^{O_\varepsilon(1)}\operatorname{poly}(N)$.}
\label{tab:comparison}
\end{table}

In pseudopolynomial time $(nT)^{O_\varepsilon(1)}\operatorname{poly}(N)$,
where $T$ is the numerical time horizon, Armbruster et al.~\cite{AGTW26}
obtain a factor $5/4+\varepsilon$ and Tong and Xu~\cite{TX26} obtain
$21/17+\varepsilon$, both for an arbitrary number of identical machines.
Our algorithm enumerates a relaxation of constant order depending on
$\varepsilon$. Writing $\bar\varepsilon=\min\{\varepsilon,1\}$,
Proposition~\ref{prop:runtime} gives the explicit implementation bound
$n^{2^{2^{O(1/\bar\varepsilon)}}}\operatorname{poly}(N)$.

\subsection{The tradeoff behind the improvement}
The block construction distinguishes a nonlocal job's two boundary blocks,
which contain its release and deadline endpoints, from its eligible interior
placements. An interior placement belongs to a superblock entirely contained
in the job's window. A local job is eligible in only one block.
The relaxation records the distribution of the feasible configuration in each
block. It also records joint distributions on every set of at most $D$ blocks,
where $D$ depends only on the accuracy. Each table assigns probability only
to tuples of configurations with no repeated job identity. All tables agree
with the single-block distributions.

A sampling scheme partitions the blocks into small, disjoint groups. We
sample a tuple from each group's table, independently between groups. A job
can appear at most once inside a group. A nonlocal job can therefore appear
in both of its boundary blocks only if those blocks belong to different groups.
If two jobs have strictly interleaving boundary pairs, the shorter job fits
the other job's intervening boundary slot. This creates a possible repair:
a missing job can use the spare boundary slot of a duplicated job.

There are two coupled obstructions to exploiting these repairs. Many missing
jobs may want the same spare slot, and the interior recovery may already use
part of a missing job's capacity. We address the first obstruction with a
capacitated flow on crossing pairs of active jobs, whose duplication
probabilities are at least a small threshold. We address the second by reserving
separate fractions of each missing job's unit capacity for interior
assignments and boundary repairs. If $a_j,b_j$ are the boundary marginals,
$h_j$ is the probability of duplication, $c_j$ is its interior mass, and
$q_j=1-a_j-b_j+h_j$ is the probability of boundary absence, the reservations
are $c_j/q_j$ and $h_j/q_j$ when $q_j>0$. They fit together because
$a_j+b_j+c_j\le 1$ implies $c_j+h_j\le q_j$. A bounded-group variance
estimate controls the mass lost when the proposed assignments exceed a
reservation or a spare slot. This yields a feasible fractional matching;
the algorithm obtains at least its value by maximum integral matching.

Now write $F_i$ for retained fractional mass, $W_i$ for expected boundary
duplication, and $H_i$ for the repair-flow value of scheme $i$. Up to accuracy
losses, its expected matching value is at least
\[
  F_i-W_i+H_i.
\]
A large flow thus improves the current output. A small flow is useful for
a different reason. A residual cut identifies boundary edges that have few
crossings among themselves; the active edges left unprotected have total
duplicate mass at most the flow value, while inactive jobs contribute only
an accuracy loss. A component-preserving separator places the
protected endpoints into new bounded-size groups, losing little fractional
mass. Sampling these groups prevents their boundary duplicates. The resulting
tradeoff is
\[
  F_{i+1}\ge F_i-O(\delta F),\qquad
  W_{i+1}\le H_i+O(\delta F),
\]
where $F$ is the initial relaxation value and $\delta$ is an internal accuracy
parameter. The same quantity $H_i$ rewards the current scheme and bounds
the next scheme's remaining duplication.

\paragraph{An accounting example.}
Ignore the accuracy losses and normalize $F$ to $12$ units. The initial
duplication bound is $W_0\le 3$. After two coordination steps,
$W_1\le H_0$ and $W_2\le H_1$. Set $H_2=0$ for the final scheme. The three scores are then bounded below by
\[
  9+H_0,\qquad 12-H_0+H_1,\qquad 12-H_1.
\]
These three lower bounds sum to $33$, regardless of the two flow values.
Hence at least one score is at least $11$, giving the retained fraction
$11/12$. For example, $H_0=2$ and $H_1=1$ make all three displayed lower
bounds equal to $11$.
This idealized example illustrates the accounting identities. The formal
proof restores the accuracy losses, computes the three scores deterministically,
and samples a scheme of maximum score.

The structural reason for taking two steps is separate from this algebra.
The separator accommodates two protected boundary graphs drawn on opposite
sides of the cyclic block order. A third graph would require a new structural
argument.

\subsection{Relation to previous techniques}
Tong and Xu~\cite[Sections~2--3]{TX26} coordinate noncrossing boundary edges
by gluing three-block distributions along a triangulation, using agreement on
shared pairs. Their conclusion explicitly proposes both larger crossing
exchanges and coordination of crossing graphs admitting bounded-size
tree-decomposition bags~\cite[Section~5]{TX26}. Thus crossing-family
coordination is an existing research direction. Our specific contribution
is a flow/residual-cut dichotomy, compatible interior and repair capacity
reservations, and a two-page separator that preserves earlier groups.
These ingredients implement two rounds of coordination with a quantitative
loss bound. We sample disjoint groups from larger tables that agree only on
single-block marginals. Each regrouping defines a fresh sampling scheme.

The structural preprocessing follows the block framework of~\cite{AGTW26}.
Its nested-scale density refinement also follows~\cite[Appendix~A]{TX26},
with scale growth chosen here to accommodate the larger table order.
Our separator uses the planar separator theorem of Lipton and
Tarjan~\cite{LT79}. The additional requirement is to discard entire earlier
components so that new groups preserve all surviving earlier coordination.

For historical context, Chuzhoy, Ostrovsky, and Rabani~\cite{COR06} obtained
a factor arbitrarily close to $e/(e-1)$ for job interval selection, whose
formulation permits explicit intervals or intervals specified by job
parameters. Im, Li, and Moseley~\cite{ILM20} obtained a retained fraction
$\alpha_0-\varepsilon$ for a constant $\alpha_0>0.6448$ on a fixed number
of machines. Under a different restriction,
Hyatt-Denesik, Rahgoshay, and Salavatipour~\cite{HRS20} give a randomized
approximation scheme when both the number of machines and the number of
distinct processing times are fixed. Our theorem allows arbitrary numbers
of machines and distinct processing times.

\paragraph{Organization.}
\Cref{sec:relaxation} defines the relaxation and its parameters.
\Cref{sec:coordination} proves the flow-to-coordination tradeoff, and
\cref{sec:matching} checks the three matching-capacity accounts.
\Cref{sec:main-proof} combines the guarantees and bounds the running time.
\Cref{sec:preprocessing,sec:separator} contain the structural proofs.
\Cref{sec:multiple-machines} gives the extension to an arbitrary number of
identical machines.

\section{Model and a local relaxation}\label{sec:relaxation}
We first consider one machine. Appendix~\ref{sec:multiple-machines} extends
the result to an arbitrary number of identical machines.
A job $j$ has an integral release time $r_j$, deadline $d_j$, and positive
integral processing time $p_j$. Its window is $[r_j,d_j)$. A schedule assigns
each selected job one half-open execution interval of length $p_j$ contained
in its window; execution intervals must be disjoint. All numbers are given
in binary. We write $n$ for the number of jobs, $N$ for the input encoding
length, and $\mathrm{OPT}$ for the maximum number of scheduled jobs.
Delete jobs with $p_j>d_j-r_j$, and solve instances with at most one job directly. A common
translation of all times lets us assume $r_j\ge0$. Put $T=\max_jd_j$.

Fix an integer $Q\ge100$ and put $\delta=1/Q$. The same integer $Q$
will count the mass portions in rounding and the grouping scales in preprocessing.
A block partition $\mathcal B$ divides $[0,T)$ into consecutive half-open
intervals. A superblock partition $\mathcal S$ groups consecutive blocks.
The release block of $j$ contains $r_j$; its deadline block $[u,v)$ is the
unique block with $d_j\in(u,v]$. A job is \emph{local} if these blocks agree.
A local job is eligible only in that block. A nonlocal job is eligible in
its two boundary blocks and in any other block whose entire superblock is
contained in its window. The latter occurrences are called interior.
An eligible schedule executes each job in one eligible block.
\begin{table}[t]
\centering
\small
\begin{tabular}{@{}p{0.25\textwidth}p{0.69\textwidth}@{}}
\toprule
Notation & Meaning \\
\midrule
$\delta$, $Q=\delta^{-1}$ & Loss parameter and integer number of portions or scales. \\
$K$, $D(K)$ & Jobs per block; maximum number of blocks in a local table. \\
$m_i$, $\beta=Km_i$ & Group-size bound at step $i$; jobs in one sampled group. \\
$F$, $F_i$ & Original LP value; retained placement mass at step $i$. \\
$a_j,b_j,c_j$ & Retained release, deadline, and interior marginals of nonlocal job $j$. \\
$h_j,q_j$ & Probabilities of two boundary copies and no boundary copy. \\
$W_i$, $H_i$ & Total duplication probability and maximum repair flow at step $i$. \\
$\mathcal J_{\rm act}$, $\mathcal U$ & Active jobs and the unprotected jobs identified by the residual cut. \\
$M_j$, $V_j$ & Indicators of no boundary copy and two boundary copies. \\
$C_{\mathrm{int}}$, $J_{\mathrm{out}}$ & Total retained interior mass and the algorithm's output job set. \\
$\theta$, $\eta$, $\kappa(m)$ & Activity threshold, middle-arc capacity, and crossing-degree bound. \\
$\tau=6Q$, $J_i$ & Refinement base and nested retained job sets in Appendix~\ref{sec:preprocessing}. \\
\bottomrule
\end{tabular}
\caption{Principal notation. The per-job quantities $a_j,b_j,c_j,h_j,q_j$ and indicators $M_j,V_j$ refer only to nonlocal jobs. All job marginals and probabilities refer to the current grouping; step indices are added when several groupings are compared. The definitions in the text specify the degenerate cases.}\label{tab:notation}
\end{table}

\begin{lemma}[Preprocessing]\label{lem:preprocessing}
For any computable integer-valued function $D(K)\ge1$ depending only on
$K$ and $\delta$, a deterministic algorithm either solves the instance
exactly or constructs $\delta^{-1}$ candidate block--superblock partitions
$(\mathcal B_\ell,\mathcal S_\ell)$, each with its own cap $K_\ell$ depending
only on $\delta$ and $D$. For at least one candidate $\ell$, writing
$K=K_\ell$, $\mathcal B=\mathcal B_\ell$, and $\mathcal S=\mathcal S_\ell$,
there is an eligible schedule $J'$ that crosses no block boundary, uses at most $K$
jobs per block, and satisfies
\begin{equation}\label{eq:preprocessing}
 |J'|\ge(1-2\delta)\mathrm{OPT},\qquad
 |\mathcal B|\le\delta|J'|,\qquad
 KD(K)|\mathcal S|\le\delta^3|J'|.
\end{equation}
The algorithm runs in $n^{O_\delta(1)}\operatorname{poly}(N)$ time, where
the hidden constant may depend on the prescribed function $D$.
\end{lemma}
The proof, including the dependence on $D$, appears in
Appendix~\ref{sec:preprocessing}. The schedule in the conclusion is an
analysis witness; the preprocessing algorithm computes the candidate
partitions. The table-order function used
by our algorithm is specified in~\eqref{eq:constants} below.

For a fixed candidate $\ell$, write $\mathcal B,\mathcal S$ for its
partitions and abbreviate $K=K_\ell$ and $D=D(K_\ell)$.
Let $\mathcal C_B$ consist of all eligible job
sets of size at most $K$ schedulable inside block $B$. Store one feasible
execution schedule for each set, including the empty set. To construct this
family, enumerate job sets and orders and schedule each order at its earliest
possible start times inside $B$. An order is feasible if and only if this
procedure completes every job by its deadline and completes all jobs by
the right endpoint of $B$. Thus the
family can be constructed in $n^{O(K)}\operatorname{poly}(N)$ time per block.
For nonempty $U\subseteq\mathcal B$ of size at most $D$, let
$\mathcal T_U$ consist of tuples $(C_B)_{B\in U}$ with $C_B\in\mathcal C_B$
and with no job identity appearing twice.

The relaxation has a probability distribution $x_{B,\cdot}$ on
$\mathcal C_B$ and a distribution $\mu_U$ on $\mathcal T_U$ for every such
$U$. It imposes
\begin{align}
 \sum_{\boldsymbol C\in\mathcal T_U:C_B=C}\mu_U(\boldsymbol C)
 &=x_{B,C} &&(B\in U,\ C\in\mathcal C_B),\label{eq:marginals}\\
 y_j:=\sum_{B\in\mathcal B}\sum_{C\in\mathcal C_B:j\in C}x_{B,C}
 &\le1 &&(j\in J),\label{eq:jobLP}
\end{align}
and maximizes $F=\sum_jy_j$. Every distribution is nonnegative and normalized.
The consistency constraints equate only the single-block marginals
in~\eqref{eq:marginals}.

Every candidate has $O(n)$ blocks, as follows from the construction in
Appendix~\ref{sec:preprocessing}. Consequently the explicit LP has
$n^{O(KD)}$ variables and constraints. An eligible schedule with cap $K$
induces a feasible integral point, using the restriction of that schedule in
every table. Solve each candidate's LP with its own parameters
$K_\ell$ and $D(K_\ell)$, obtaining value $F_\ell$. Discard candidate $\ell$
if $|\mathcal B_\ell|>\delta F_\ell$ or
$K_\ell D(K_\ell)|\mathcal S_\ell|>\delta^3F_\ell$, and choose a remaining
candidate with largest value. From now on $\mathcal B,\mathcal S,K,D,F$
denote the selected partitions, parameters, and LP value.
Lemma~\ref{lem:preprocessing} ensures that at least one candidate survives
and that the selected solution satisfies
\begin{equation}\label{eq:LPbounds}
 F\ge(1-2\delta)\mathrm{OPT},\qquad
 |\mathcal B|\le\delta F,\qquad KD|\mathcal S|\le\delta^3F.
\end{equation}
From this point onward, $F$ denotes this fixed original LP value.

\section{Coordinating the boundary blocks}\label{sec:coordination}
A grouping $\mathcal Q$ partitions a subset of the original blocks into
sets of size at most $m\le D$. Deleted blocks are not sampled. Independently
for each $G\in\mathcal Q$, sample a tuple from $\mu_G$ and use the stored
execution intervals of its configurations as candidate slots. The matching
step will choose the jobs actually executed. Groups need not be consecutive.
Put $\beta =Km$; any sampled group contains at most $\beta $ distinct jobs.
Write $F'$ for the sum of retained placement marginals.

For a nonlocal job $j$, let $a_j,b_j$ be its marginals in its retained release
and deadline blocks, respectively, taking a marginal to be zero if the
corresponding block was deleted. Let $c_j$ be its total retained interior
mass. Define
\begin{equation}\label{eq:hq}
 \begin{aligned}
 h_j&=\begin{cases}
 a_jb_j,&\text{if both boundary blocks survive in distinct groups},\\
 0,&\text{otherwise},
 \end{cases}\\
 q_j&=1-a_j-b_j+h_j,\qquad W=\sum_jh_j.
 \end{aligned}
\end{equation}
The table for a group never repeats a job; different groups are sampled
independently. Hence $h_j$ is the probability that both boundaries contain
$j$, and $q_j$ is the probability that neither does. In particular,
\begin{equation}\label{eq:reservation}
 c_j+h_j\le q_j,\qquad W\le F'/4\le F/4.
\end{equation}
The first inequality is equivalent to $a_j+b_j+c_j\le1$. For the second,
$h_j\le(a_j+b_j)^2/4\le(a_j+b_j)/4$, since $a_j+b_j\le1$.
All sums over boundary pairs below are over nonlocal jobs.

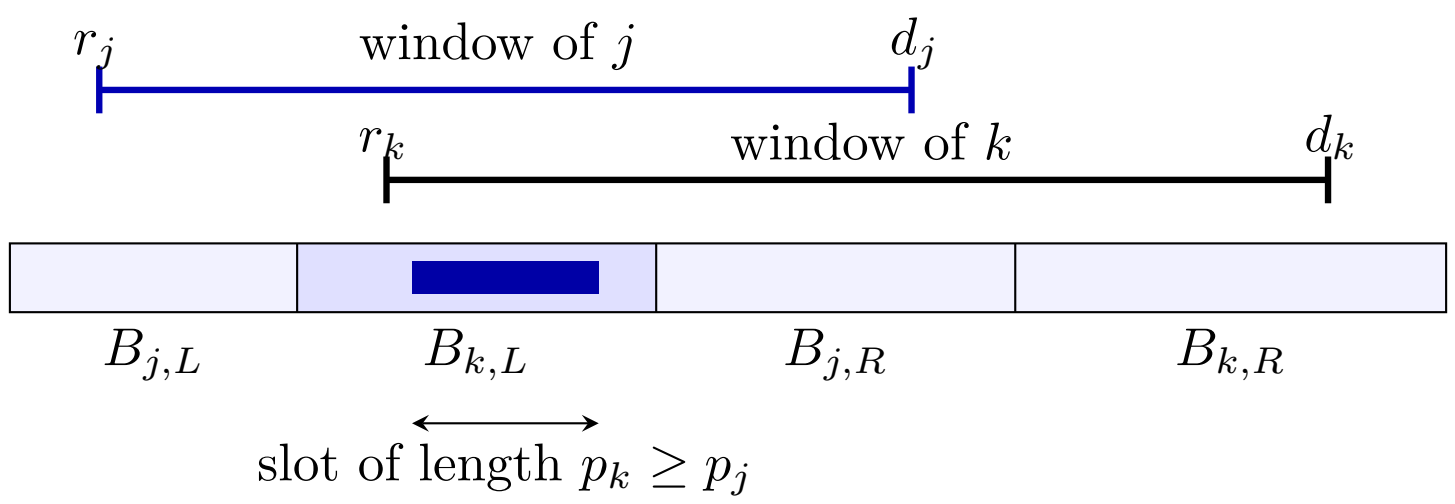
\begin{figure}[t]
\centering
\begin{tikzpicture}[x=0.95cm,y=0.7cm,>=stealth,font=\small]
  \fill[blue!5] (0,0) rectangle (2,0.65);
  \fill[blue!12] (2,0) rectangle (4.5,0.65);
  \fill[blue!5] (4.5,0) rectangle (7,0.65);
  \fill[blue!5] (7,0) rectangle (10,0.65);
  \draw (0,0) rectangle (10,0.65);
  \foreach \x in {2,4.5,7} {\draw (\x,0)--(\x,0.65);}
  \node at (1,-0.4) {$B_{j,L}$};
  \node at (3.25,-0.4) {$B_{k,L}$};
  \node at (5.75,-0.4) {$B_{j,R}$};
  \node at (8.5,-0.4) {$B_{k,R}$};
  \draw[very thick,blue!70!black,|-|] (0.6,2.1)--(6.3,2.1);
  \node[above] at (0.6,2.1) {$r_j$};
  \node[above] at (6.3,2.1) {$d_j$};
  \node[above] at (3.4,2.1) {window of $j$};
  \draw[very thick,|-|] (2.6,1.25)--(9.2,1.25);
  \node[above] at (2.6,1.25) {$r_k$};
  \node[above] at (9.2,1.25) {$d_k$};
  \node[above] at (6.0,1.25) {window of $k$};
  \fill[blue!65!black] (2.8,0.17) rectangle (4.1,0.48);
  \draw[<->] (2.8,-1.05)--(4.1,-1.05);
  \node[below] at (3.45,-1.05) {slot of length $p_k\geq p_j$};
\end{tikzpicture}
\caption{One of the two strictly interleaving orders. The whole block $B_{k,L}$ lies in the window of $j$. A slot generated by $k$ in this block can therefore serve $j$ when $p_j\leq p_k$. The opposite interleaving order uses $B_{k,R}$. The drawing is schematic.}\label{fig:crossing}
\end{figure}

Set $\theta=\delta^2$ and $\eta=\delta^6/\beta $. Let $\mathcal J_{\mathrm{act}}=\{j:h_j\ge\theta\}$ be the set of active jobs. Two boundary pairs cross when their four blocks strictly
interleave in chronological order. Orient a crossing $j\to k$ if
$p_j\le p_k$, with any fixed tie-breaking order on job identities. One
boundary block of $k$ lies strictly between the two boundary blocks of $j$,
and is therefore entirely contained in $[r_j,d_j)$. The execution slot of
$k$ in that block can accommodate $j$. A crossing is permitted if both jobs
are active and the four boundary blocks belong to four distinct groups.
See Figure~\ref{fig:crossing}.

The flow below measures repairs available to the current grouping. A large
flow improves its matching guarantee; a small flow allows the residual-cut
and separator construction to reduce duplication in the next grouping.
Comparing three groupings uses this tradeoff twice, before any sampling.

Build a directed network with source $s$, sink $t$, left and right copies
$j_L,j_R$ of each active job, arcs $s\to j_L$ and $j_R\to t$ of capacity
$h_j$, and an arc $j_L\to k_R$ of capacity $\eta$ for each permitted
crossing $j\to k$. Let $u_{jk}$ be a maximum flow on the middle arcs and
$H$ its value. In particular,
\begin{equation}\label{eq:flow}
 0\le u_{jk}\le\eta,\qquad
 \sum_ku_{jk}\le h_j,\qquad
 \sum_ju_{jk}\le h_k,\qquad H=\sum_{jk}u_{jk}.
\end{equation}
Let $X$ and $Y$ be the sets of reachable left and right copies, respectively,
in the residual graph from $s$. Define
\[
 \mathcal U=\{j\in\mathcal J_{\mathrm{act}}:j_L\notin X\text{ or }j_R\in Y\}.
\]
Protect every job in $\mathcal J_{\mathrm{act}}\setminus\mathcal U$. On the retained block
vertices draw its boundary pair as a chord, merging parallel chords.
Call the resulting simple graph the protected graph.

We use the following separator statement, proved in
Appendix~\ref{sec:separator}. For two simple graphs on the same cyclically
ordered $b$ vertices, each with at most $\kappa$ crossings per edge, suppose the
first graph has components of size at most $m$. For any integer $M\ge1$,
we can delete whole components of the first graph, containing at most
\begin{equation}\label{eq:separator-bound}
 \frac{140c(\kappa)mb}{\sqrt M}\quad\text{vertices},\qquad
 c(\kappa)=1+2\kappa(\kappa+1),
\end{equation}
so that every remaining component of their union has size at most $M$.
Crossings here refer to chords with strictly interleaved endpoints; edges
sharing an endpoint do not cross.

Define the constants
\begin{equation}\label{eq:constants}
 \kappa(m)=4Km\delta^{-6},\qquad m_0=1,\qquad
 m_{i+1}=[280Km_i c(\kappa(m_i))]^2\ (i=0,1),\qquad D(K)=m_2.
\end{equation}
They are integers because $\delta^{-1}$ is an integer. This is the function
$D$ used in preprocessing and in the LP.

Initially every block is a singleton group, and the old graph is edgeless.
At step $i\in\{0,1\}$, compute the network and protected graph for the
current grouping. Apply~\eqref{eq:separator-bound} to the old graph and
protected graph with $m=m_i$, $\kappa=\kappa(m_i)$, and $M=m_{i+1}$. Delete the whole
old components returned by the procedure and group the remaining blocks
by connected components of the union. The first resulting old graph is
the retained protected graph. After the second step no further old graph
is needed. Write $F_i,W_i,H_i$ for the retained mass, duplication mass,
and maximum flow at step $i$, and put $H_2=0$.

\begin{lemma}[Coordination]\label{lem:coordination}
Each of the two steps is well-defined and satisfies
\begin{equation}\label{eq:coordination}
 F_{i+1}\ge F_i-\delta F/2,\qquad
 W_{i+1}\le H_i+\delta F/2.
\end{equation}
\end{lemma}
\begin{proof}
Fix one of the two steps and its current grouping. We first bound the mass
that cannot be protected, then establish the separator hypotheses, and
finally track the retained marginals after regrouping.

\paragraph{The residual cut charges unprotected mass to the flow.}
The sink is not reachable from $s$ in the residual network of a maximum
flow. If $j_L\notin X$, then $s\to j_L$ has no positive residual
capacity; otherwise that arc would make $j_L$ reachable. Its flow is
therefore $h_j$. Similarly, $j_R\in Y$ implies that $j_R\to t$ is
saturated, since a positive residual capacity would make $t$ reachable.

These two saturation statements alone would allow a factor-two bound.
The following disjointness avoids that loss. If $j_L\notin X$ and
$k_R\in Y$, then $u_{jk}=0$: positive flow would create the reverse
residual arc $k_R\to j_L$, making $j_L$ reachable. Consequently,
\begin{align*}
 \sum_{\substack{j\in\mathcal J_{\mathrm{act}}:\,j_L\notin X}}h_j
 &=\sum_{\substack{j,k:\,j_L\notin X}}u_{jk},\\
 \sum_{\substack{k\in\mathcal J_{\mathrm{act}}:\,k_R\in Y}}h_k
 &=\sum_{\substack{j,k:\,k_R\in Y}}u_{jk}
\end{align*}
count disjoint sets of positive middle-arc flows. A job in both defining
classes of $\mathcal U$ is counted twice on the left below, which only
increases the bound. Thus
\begin{equation}\label{eq:unprotected}
 \sum_{j\in\mathcal U}h_j
 \le \sum_{\substack{j\in\mathcal J_{\mathrm{act}}:\,j_L\notin X}}h_j
    +\sum_{\substack{j\in\mathcal J_{\mathrm{act}}:\,j_R\in Y}}h_j
 \le H_i.
\end{equation}
This is the source of the large-flow/small-duplication tradeoff: protecting
all other active pairs leaves at most $H_i$ active duplicate mass unpaid.

\paragraph{Protected edges have bounded crossing degree.}
For protected jobs $j,k$, a permitted arc $j_L\to k_R$ goes from a
reachable vertex to an unreachable one. It must be saturated, so
$u_{jk}=\eta$. The total flow incident to the left and right copies of
one job is at most $2h_j\le2$. There are therefore at most $2/\eta$
permitted crossing arcs incident to a fixed protected job.

If two protected pairs cross but the crossing is not permitted, their
boundary blocks use fewer than four groups. Each pair individually uses
two distinct groups because it is active. Hence the other pair has a
boundary in one of the fixed pair's two groups. Activity gives
$a_j,b_j\ge\theta$: their product is at least $\theta$, and neither
factor exceeds one. A group has expected total placement count at most
$\beta=Km$. At most $\beta/\theta$ active jobs can consequently have a
boundary in any one group. The nonpermitted crossings of the fixed pair
number at most $2\beta/\theta$. Combining the two cases yields
\[
 \frac2\eta+\frac{2\beta}{\theta}
 \le4\beta\delta^{-6}=\kappa(m_i).
\]
For a simple protected edge, fix any job represented by that edge and
apply the preceding count. Every distinct crossing simple edge has a
represented crossing job, so merging parallel edges cannot increase the
bound.

\paragraph{The separator applies at both steps.}
Initially the old graph has no edges. After the first step, it is a
restriction of the first protected graph, and its connected components
are the groups of size at most $m_1$. Its crossing degree is at most
$\kappa(1)\le\kappa(m_1)$. The second protected graph has crossing degree
at most $\kappa(m_1)$ by the preceding argument. These are exactly the
two applications for which the hypotheses of
Lemma~\ref{lem:two-page} have been established.

At step $i$, the separator removes at most
$140c(\kappa(m_i))m_i|\mathcal B|/\sqrt{m_{i+1}}$ blocks; we used the
original number of blocks to upper-bound the current number. Each block
has placement mass at most $K$, since every one of its configurations
has at most $K$ jobs. The choice of $m_{i+1}$ gives
\[
 K\frac{140c(\kappa(m_i))m_i|\mathcal B|}{\sqrt{m_{i+1}}}
 =\frac{|\mathcal B|}{2}\le\frac{\delta F}{2}.
\]
This proves the retained-mass inequality.

\paragraph{Regrouping eliminates the protected duplicates.}
Both endpoints of every surviving protected edge lie in one new connected
component. Its job cannot be duplicated when that component is sampled
from its table. If either endpoint was deleted, its new duplication
probability is also zero. A pair already internal to an old group remains
internal or loses both endpoints, because we delete whole old groups and
never split a surviving one.

For any other pair, retaining both endpoints preserves its two individual
marginals. Merging its groups changes its duplication probability from
$a_jb_j$ to zero; deleting an endpoint also changes it to zero. Thus no
pair's duplication probability increases. Active unprotected pairs
contribute at most $H_i$ by~\eqref{eq:unprotected}. For an inactive pair,
$h_j<\delta^2$ and $2\sqrt{h_j}\le a_j+b_j$, including the case $h_j=0$.
Hence $h_j\le\delta(a_j+b_j)/2$. Summing over those pairs gives at most
$\delta F_i/2\le\delta F/2$. This proves the duplicate-mass inequality.
\end{proof}

\subsection{The two-step graph invariant}\label{sec:two-round-limit}
The two coordination steps use the following graph invariant. After one step, the old graph is a single protected
chord family with bounded crossing degree. The second step places that
family and the new protected family on the two hemispheres used in
Lemma~\ref{lem:two-page}. After this step, the old graph would be their
\emph{union}. The established crossing bounds apply separately to its two
families. A further application of the separator would require a crossing
bound for their union in the fixed cyclic order.

This distinction persists even when all old components are tiny. For any
$t\ge1$, put vertices in cyclic order
\[
 x,a_1,\ldots,a_t,y,b_t,\ldots,b_1.
\]
The family consisting of $xy$ and the family consisting of the nested
edges $a_ib_i$ are each crossing-free. Their union is a matching, so each
component has two vertices, but $xy$ crosses all $t$ edges of the other
family. Thus neither pagewise crossing bounds nor a bound on old-component
size supplies the one-page hypothesis needed for another invocation of
our lemma. The example isolates the failure of this structural implication.
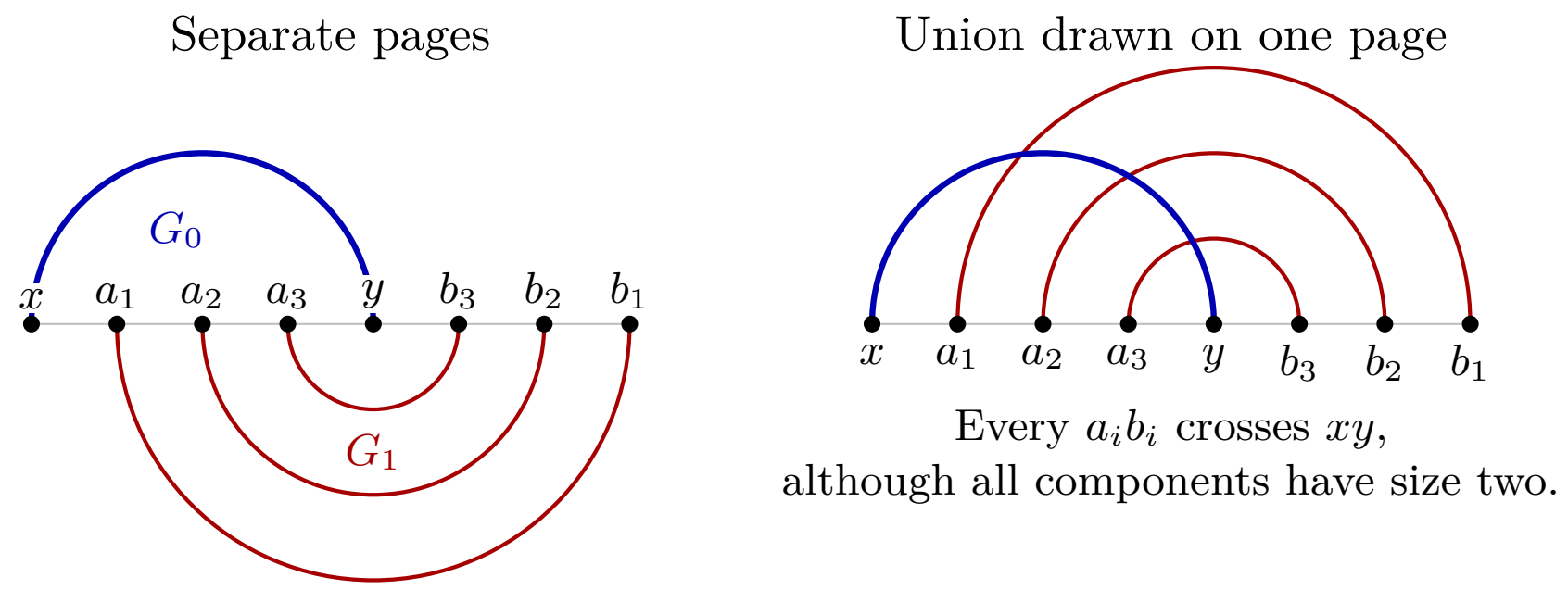
\begin{figure}[t]
\centering
\begin{tikzpicture}[x=0.61cm,y=0.61cm,font=\footnotesize]
\begin{scope}
  \node[font=\small] at (3.5,3.35) {Separate pages};
  \draw[gray!50] (0,0)--(7,0);
  \draw[very thick,blue!70!black] (0,0) arc[start angle=180,end angle=0,radius=2];
  \foreach \a/\b/\r in {1/7/3,2/6/2,3/5/1} {
    \draw[thick,red!65!black] (\a,0) arc[start angle=180,end angle=360,radius=\r];
  }
  \foreach \x/\lab in {0/x,1/a_1,2/a_2,3/a_3,4/y,5/b_3,6/b_2,7/b_1} {
    \fill (\x,0) circle[radius=1.7pt];
    \node[fill=white,inner sep=1pt,above=2pt] at (\x,0) {$\lab$};
  }
  \node[blue!70!black] at (1.7,1.1) {$G_0$};
  \node[red!65!black] at (4,-1.5) {$G_1$};
\end{scope}
\begin{scope}[xshift=6cm]
  \node[font=\small] at (3.5,3.35) {Union drawn on one page};
  \draw[gray!50] (0,0)--(7,0);
  \foreach \a/\b/\r in {1/7/3,2/6/2,3/5/1} {
    \draw[thick,red!65!black] (\a,0) arc[start angle=180,end angle=0,radius=\r];
  }
  \draw[very thick,blue!70!black] (0,0) arc[start angle=180,end angle=0,radius=2];
  \foreach \x/\lab in {0/x,1/a_1,2/a_2,3/a_3,4/y,5/b_3,6/b_2,7/b_1} {
    \fill (\x,0) circle[radius=1.7pt];
    \node[fill=white,inner sep=1pt,below=3pt] at (\x,0) {$\lab$};
  }
  \node[align=center] at (3.5,-1.55) {Every $a_i b_i$ crosses $xy$,\\although all components have size two.};
\end{scope}
\end{tikzpicture}
\caption{Why the established pagewise bounds do not automatically apply to the accumulated old graph. In the order $x,a_1,\ldots,a_t,y,b_t,\ldots,b_1$, the two matchings $G_0=\{xy\}$ and $G_1=\{a_ib_i:1\le i\le t\}$ are each crossing-free. Their union has only two-vertex components, but $xy$ crosses all $t$ other edges in a common one-page drawing. The case $t=3$ is shown. This is a counterexample to an inference from pagewise crossing bounds and component size alone.}\label{fig:two-page-limit}
\end{figure}

To separate the algebraic opportunity from this missing structural step,
suppose hypothetically that one could carry out $t$ steps with the same
bounds as Lemma~\ref{lem:coordination}, and apply Lemma~\ref{lem:rounding}
to all $t+1$ groupings. Put $H_t=0$. The same calculation would give
\begin{align*}
 \sum_{i=0}^{t}(F_i-W_i+H_i)
 &\ge \sum_{i=0}^{t}F_i-W_0-\frac{t\delta F}{2},\\
 \mathbb E|J_{\mathrm{out}}|
 &\ge \left(1-\frac{1}{4(t+1)}
 -\delta\left(4+\frac t4+\frac{t}{2(t+1)}\right)\right)F.
\end{align*}
Here $F_i\ge F-i\delta F/2$ and $W_0\le F/4$ explain every term in the
second line. Under these hypotheses, extending the structural argument to
more rounds would therefore improve the approximation ratio.

\section{Matching sampled slots}\label{sec:matching}
For any grouping, treat every sampled execution as a slot $[s,t)$.
A job $j$ is compatible with the slot if
\begin{equation}\label{eq:compatibility}
 \min\{t,d_j\}-\max\{s,r_j\}\ge p_j.
\end{equation}
Compute a maximum bipartite matching between all jobs and all slots. An
assigned job starts at $\max\{s,r_j\}$. Since the slots are disjoint and
matching uses each job and slot at most once, this is a feasible schedule.
The fractional matching used next is only an analysis certificate.

\begin{lemma}[Rounding]\label{lem:rounding}
For any grouping with group size at most $m\le D$, and any nonnegative
weights on permitted crossings satisfying~\eqref{eq:flow}, the expected
maximum matching cardinality is at least
\[
 F'-W+H-4\delta F.
\]
\end{lemma}

\subsection{A variance bound}
The following fact allows arbitrary correlations inside one sampled group.
\begin{lemma}\label{lem:variance}
Suppose independent random group samples each contain at most $\beta $ job
occurrences, with no repeated identity. The occurrence record of a job in a
group specifies the block
containing it, or records its absence. For each job $j$, let
$T_j\in\{0,1\}$ be determined only by its occurrence records in a fixed
set of at most two groups. If
$0\le w_j\le a$, then
\[
 \operatorname{Var}\!\left(\sum_jw_jT_j\right)
 \le4\beta a\,\mathbb E\!\left[\sum_jw_jT_j\right].
\]
\end{lemma}
\begin{proof}
Write $Z=\sum_jw_jT_j$. If $0<\Pr(T_j=1)<1$, couple the conditional
experiments given $T_j=1$ and $T_j=0$ by using identical samples for all
groups outside the at most two determining groups. Independence makes
this coupling valid: these outside groups retain their original joint
law under either condition. The two samples in the determining groups
contain altogether at most $4\beta $ job identities. A flag can change only
if its job is among these identities, including the case where the job
appears in both samples but in different blocks. Therefore the coupled values of
$Z$ differ in absolute value by at most $4\beta a$, and
\[
 \operatorname{Cov}(T_j,Z)
 =\Pr(T_j=1)\Pr(T_j=0)
   \bigl(\mathbb E[Z\mid T_j=1]-\mathbb E[Z\mid T_j=0]\bigr)
 \le4\beta a\,\mathbb E T_j.
\]
If $T_j$ is constant, its covariance is zero. Finally sum these inequalities
with weights $w_j$, using
$\operatorname{Var}(Z)=\sum_jw_j\operatorname{Cov}(T_j,Z)$.
\end{proof}

\subsection{Constructing the fractional matching}
\begin{proof}[Proof of Lemma~\ref{lem:rounding}]
Let $M_j$ and $V_j$ indicate boundary absence and boundary duplication.
Then $\mathbb EM_j=q_j$ and $\mathbb EV_j=h_j$. The baseline counts each
sampled local job and each boundary-present nonlocal job exactly once.
Writing $C_{\mathrm{int}}=\sum_jc_j$, its expected size is $F'-C_{\mathrm{int}}-W$.
Its actual assignment to boundary slots will be completed after repairs.
For each absent nonlocal job reserve capacity $c_j/q_j$ for interior
assignments and $h_j/q_j$ for boundary repairs. Their sum is at most one
by~\eqref{eq:reservation}. Whenever $q_j=0$, both numerators are zero;
such terms are defined to be zero.
Table~\ref{tab:capacity-ledger} records the capacities that will be checked
jointly after the two kinds of assignment have been constructed.
Section~\ref{sec:capacity-example} illustrates these reservations on a
concrete feasible set of local tables.
\begin{table}[!ht]
\centering
\small
\renewcommand{\arraystretch}{1.12}
\begin{tabular}{@{}p{0.23\textwidth}p{0.31\textwidth}p{0.39\textwidth}@{}}
\toprule
Account & Assigned mass & Capacity check \\
\midrule
Absent recipient $j$ & Interior at most $(c_j/q_j)M_j$; repairs at most $(h_j/q_j)M_j$ & Sum at most $M_j$, since $c_j+h_j\le q_j$. \\
Duplicated donor $k$ & Repairs export at most $V_k$ in total from its two boundary slots & If $V_k=1$, two unit slots retain at least one unit for $k$. Each slot exports at most one. \\
Other baseline jobs & One unit for each sampled local job or job with exactly one boundary copy & Its own sampled slot supplies that unit. \\
Interior slots & Used only for absent recipients & The portion assignment uses at most one unit per slot. \\
\bottomrule
\end{tabular}
\caption{The simultaneous capacity accounts in the fractional-matching certificate. An absent job cannot also be a donor or a boundary-present baseline job. Ratios are used only when their denominator is positive; a zero-probability event contributes no assignment.}\label{tab:capacity-ledger}
\end{table}

\paragraph{Interior assignments.}
Fix a superblock with total interior mass $C>0$. Let $c_{j,S}$ be the
retained interior mass of $j$ in this superblock. Sort these masses by
nonincreasing processing time, breaking ties arbitrarily, and split the
resulting mass interval into $Q$ consecutive portions, each
of mass $z_S=\delta C$. A job's mass may straddle a cut. Let $c_{j,g}$ be
its mass in portion $g$. Define the demand
\[
 N_g=\sum_j(c_{j,g}/q_j)M_j.
\]
For every sampled interior slot generated by $j$ in this superblock,
allocate the fraction $c_{j,g}/c_{j,S}$ to supply for portion $g$;
let the resulting total supply be $U_g$. Terms with $c_{j,S}=0$ are absent.
The fractions over all portions sum to one for every such slot.
Both $N_g$ and $U_g$ have expectation $z_S$, by the preserved one-block
marginals. Each demand coefficient is at most one, so
Lemma~\ref{lem:variance} gives $\operatorname{Var}(N_g)\le4\beta z_S$.
Write $U_g$ as a sum of independent group contributions in $[0,\beta ]$.
For a variable $V\in[0,\beta ]$, $\operatorname{Var}(V)\le\mathbb EV^2\le
\beta \mathbb EV$, hence $\operatorname{Var}(U_g)\le \beta z_S$.

For $g\ge2$, every demanded job in portion $g$ has processing time at
most the length of every slot supplying portion $g-1$. Every demanded job's
window contains the entire superblock. Thus the corresponding fractional
bipartite graph is complete, and it can carry a flow of value
$\min\{N_g,U_{g-1}\}$. The following estimate holds for arbitrary dependence between demand and supply:
\begin{align*}
 \mathbb E\min\{N_g,U_{g-1}\}
 &=z_S-\tfrac12\mathbb E|N_g-U_{g-1}|\\
 &\ge z_S-\tfrac12\bigl(\sqrt{\operatorname{Var}(N_g)}+
                 \sqrt{\operatorname{Var}(U_{g-1})}\bigr)
 \ge z_S-\tfrac32\sqrt{\beta z_S}.
\end{align*}
Summing $g=2,\ldots,Q$ and using
$\tfrac32\sqrt{\beta C/\delta}\le\delta C+9\beta /(16\delta^2)$ gives an expected contribution
of at least $(1-2\delta)C-9\beta /(16\delta^2)$. A zero-mass superblock needs no
assignment and satisfies the same lower bound. Summing over superblocks,
and using $\beta \le KD$ and~\eqref{eq:LPbounds}, the expected interior
contribution is at least
\begin{equation}\label{eq:interior-value}
 C_{\mathrm{int}}-2\delta C_{\mathrm{int}}-
 \frac{9\beta |\mathcal S|}{16\delta^2}
 \ge C_{\mathrm{int}}-\frac{41}{16}\delta F
 \ge C_{\mathrm{int}}-3\delta F.
\end{equation}
Each supply portion is used at most once. The total demand assigned to
job $j$ over all superblocks is at most $(c_j/q_j)M_j$, so all interior
assignments respect the reserved capacities.

\paragraph{Boundary repairs.}
For each positive-weight permitted pair propose a transfer
\[
 P_{jk}=\frac{u_{jk}}{q_jh_k}M_jV_k.
\]
Positive weights imply $q_j\ge h_j\ge\theta$ and $h_k\ge\theta$.
The four determining groups are distinct, so $M_j$ and $V_k$ are
independent and $\mathbb EP_{jk}=u_{jk}$. Let
$X_j=\sum_kP_{jk}$ and $Y_k=\sum_jP_{jk}$. Scale any overloaded row
down to capacity $(h_j/q_j)M_j$, and then scale any overloaded column
down to capacity $V_k$. Column scaling cannot increase a row sum.
The total removed mass is at most the sum of the original row and
column excesses.

Put $u_j^+=\sum_ku_{jk}$ and $u_k^-=\sum_ju_{jk}$. Conditional on
$M_j=1$, every flag $V_k$ in row $j$ depends only on groups outside
$j$'s two boundary groups. Its joint distribution with the other row
flags is consequently unchanged by this conditioning. Applying
Lemma~\ref{lem:variance} with weights $u_{jk}/h_k\le\eta/\theta$ gives
\[
 \mathbb E[X_j\mid M_j=1]=u_j^+/q_j\le h_j/q_j,\qquad
 \operatorname{Var}(X_j\mid M_j=1)
 \le\frac{4\beta \eta u_j^+}{q_j^2\theta}.
\]
The same argument for a column, conditioning on $V_k=1$ and using
weights $u_{jk}/q_j\le\eta/\theta$, gives
\[
 \mathbb E[Y_k\mid V_k=1]=u_k^-/h_k\le1,\qquad
 \operatorname{Var}(Y_k\mid V_k=1)
 \le\frac{4\beta \eta u_k^-}{h_k^2\theta}.
\]
If $\mathbb EZ\le t$, then
$\mathbb E(Z-t)_+\le\mathbb E|Z-\mathbb EZ|\le
\sqrt{\operatorname{Var}(Z)}$. Multiplying the conditional bounds by
$q_j$ and $h_k$, respectively, bounds expected row and column losses by
$2\sqrt{\beta \eta u_j^+/\theta}$ and
$2\sqrt{\beta \eta u_k^-/\theta}$.
There are at most $F/(4\theta)$ active jobs, by~\eqref{eq:reservation},
and $H\le W\le F/4$. Cauchy--Schwarz therefore bounds total expected
loss by
\begin{equation}\label{eq:repair-loss}
 4\sqrt{\beta \eta/\theta}\sqrt{\frac{F}{4\theta}H}
 \le\frac{F\sqrt{\beta \eta}}{\theta}=\delta F.
\end{equation}
This also covers $H=0$, when every transfer is zero.

\paragraph{Checking all capacities simultaneously.}
Realize each surviving transfer on the donor's boundary slot lying inside
the recipient's window. The slot is compatible by the crossing geometry
and $p_j\le p_k$. Fix a duplicated donor $k$. Column scaling bounds the
\emph{total} mass exported from its two slots by one; hence the exported
mass on either individual slot is also at most one. If these two exported
masses are $e_1,e_2$, their unused capacities satisfy
$(1-e_1)+(1-e_2)\ge1$. Assign one unit of $k$ across these unused capacities,
for example $1-e_1$ on the first slot and $e_1$ on the second. This is
feasible because $e_1\le1-e_2$. The job $k$ is boundary-present and
therefore receives no interior or repair assignment of its own.

A job with exactly one boundary occurrence receives one unit on that
slot and exports nothing, since its duplication flag is zero. A sampled
local job also receives one unit on its own slot; it has no boundary-repair
or interior role. Thus all baseline jobs receive exactly one unit. The
remaining recipients are boundary-absent nonlocal jobs. Their combined
interior and repair assignment is at most
$(c_j/q_j)M_j+(h_j/q_j)M_j\le M_j$, and none is a baseline job or a donor.
Interior supplies use only sampled interior slots, whose total assigned
portion fractions are at most one. Boundary assignments use only boundary
slots, and local assignments use only local slots. These three classes of
slots are disjoint even when their generating identities coincide.
Consequently every job and every slot has total incident assignment at
most one: all assignments together form a feasible fractional matching.
Its expected value is at least
\[
 (F'-C_{\mathrm{int}}-W)+(C_{\mathrm{int}}-3\delta F)+(H-\delta F)
 =F'-W+H-4\delta F.
\]
The bipartite matching polytope is integral~\cite[Corollary~3.7a]{SC17},
so for each outcome the maximum integral matching has at least the value
of this fractional matching.
\end{proof}

\subsection{A small capacity example}\label{sec:capacity-example}
The following feasible local-table example illustrates the reservations.
Take five blocks $B_i=[4(i-1),4i)$, $i=1,\ldots,5$, each forming its own
superblock and sampling group, and two jobs
\[
 (r_j,d_j,p_j)=(1,15,1),\qquad
 (r_k,d_k,p_k)=(5,19,2).
\]
Consider a distribution that places $j$ in $B_1,B_3,B_4$ with probabilities
$1/4,1/2,1/4$, respectively, and independently places $k$ in $B_2$ or
$B_5$, each with probability $1/2$. Use execution intervals $[1,2)$,
$[8,9)$, $[12,13)$ for the three placements of $j$, and $[5,7)$,
$[16,18)$ for those of $k$. Every outcome is a feasible schedule.
Its restrictions therefore define feasible local tables of any order.

Sample the singleton groups independently from these table marginals.
For $j$, the boundary and interior masses are
$a_j=b_j=1/4$ and $c_j=1/2$, so
\[
 h_j=1/16,\qquad q_j=9/16,\qquad
 c_j/q_j=8/9,\qquad h_j/q_j=1/9.
\]
A boundary-absent job $j$ can thus receive up to $8/9$ of a unit through
interior slots and up to $1/9$ of a unit through repairs, for a total of at
most one unit. For $k$, duplication
has probability $h_k=1/4$. Its $B_2$ slot can accommodate $j$, since the
boundary pairs $B_1,B_4$ and $B_2,B_5$ cross and $p_j\le p_k$.
For the permitted flow weight $u_{jk}=\eta$, the proposed transfer on
the event $M_j=V_k=1$ is $64\eta/9$. This is at most $1/9$, since
$\eta=\delta^6/\beta\le1/64$, and so neither row nor column scaling is
needed in this example. Its expectation is exactly $\eta$. On that
event the duplicated $k$ keeps one unit in its $B_5$ slot, while the
repair uses at most one unit of its $B_2$ slot. This displays explicitly
why a donor can preserve its own baseline assignment while an absent
recipient combines two separately reserved sources of mass.

\section{The algorithm and its guarantee}\label{sec:main-proof}
Compute the two coordination steps and the three scores
$s_i=F_i-W_i+H_i$ for $i=0,1,2$, with $H_2=0$. Choose a grouping with maximum score,
sample its groups independently from their local tables, and return a
maximum compatible job--slot matching. Let $J_{\mathrm{out}}$ be the set of jobs in this
matching. Only the selected grouping is sampled.

\begin{proof}[Proof of Theorem~\ref{thm:main} for one machine]
If preprocessing solves the instance exactly, there is nothing to prove.
Otherwise $F_0=F$ and $W_0\le F/4$. Lemma~\ref{lem:coordination} gives
\begin{align*}
 \sum_{i=0}^2s_i
 &=F_0+F_1+F_2-W_0+(H_0-W_1)+(H_1-W_2)\\
 &\ge F_0+F_1+F_2-W_0-\delta F\\
 &\ge3F-W_0-\tfrac52\delta F.
\end{align*}
By Lemma~\ref{lem:rounding}, choosing a grouping with maximum score yields
\[
 \mathbb E|J_{\mathrm{out}}|\ge F-W_0/3-\tfrac{29}{6}\delta F
 \ge(11/12-29\delta/6)F
 \ge(11/12-7\delta)\mathrm{OPT}.
\]
For the last inequality, the coefficient of $F$ is positive and
\eqref{eq:LPbounds} applies; expanding the product loses at most
$(29/6+11/6)\delta=(20/3)\delta<7\delta$.

It remains to account for finite random bits and binary arithmetic.
Set $\bar\varepsilon=\min\{\varepsilon,1\}$ and
$\delta=1/\lceil100+1000/\bar\varepsilon\rceil$. All caps, refinement depths, group
sizes, and table orders are constants depending only on $\varepsilon$.
Greedy earliest-completion computations and feasibility checks use
addition, maximum, and comparison of binary endpoints. Their intermediate
numbers have encoding length polynomial in $N$ and $\log n$; no algorithm
enumerates the numerical time horizon. The explicit rational LP has
polynomial size for fixed $\varepsilon$, and a polynomial-time rational
LP algorithm produces an optimal solution of polynomial encoding
length~\cite[Section~6]{Kha80}.
All subsequent marginals, capacities, scores, maximum flows, graph
separators, and maximum matchings have polynomial encoding size and can
be computed in polynomial time in this explicit representation; for
maximum flow and matching, see~\cite[Chapters~3--4]{SC17}.

For completeness, suppose the selected grouping requires $R_s\ge1$
independent draws and each table has support at most $S_s\ge1$. Put
$b=\lceil\log_2(nR_sS_s/\delta)\rceil$. Round each table probability
down to a multiple of $2^{-b}$ and assign the remaining probability to
one outcome that originally had positive probability. The resulting
distribution has total-variation distance at most $S_s2^{-b}$ from the
original distribution. The product distributions
have distance at most $R_sS_s2^{-b}\le\delta/n$, by coupling the draws
independently. Output cardinality lies in $[0,n]$, so expected cardinality
falls by at most $\delta$. Sampling each rounded table uses exactly $b$
random bits and preserves feasibility. If there are no retained groups,
no draw is required; the score bound already handles this case.
Because the nonempty instance has $\mathrm{OPT}\ge1$, the implemented
algorithm retains at least $(11/12-8\delta)\mathrm{OPT}$ in expectation.
Finally,
\[
 \frac{11}{12}-\frac1{12/11+\varepsilon}
 =\frac{121\varepsilon}{12(12+11\varepsilon)}
 \ge\frac{121\bar\varepsilon}{276}>8\delta.
\]
The algorithm therefore has the claimed approximation factor and running
time $n^{O_\varepsilon(1)}\operatorname{poly}(N)$.
\end{proof}

\subsection{Dependence on the accuracy parameter}
We bound the running time of the explicit enumeration above, including
the nested scales in Appendix~\ref{sec:preprocessing}.

\begin{proposition}\label{prop:runtime}
With the least-multiple choice of grouping scales in the preprocessing
construction and $\bar\varepsilon=\min\{\varepsilon,1\}$, the one-machine
algorithm can be implemented in
\[
 n^{\,2^{\,2^{O(1/\bar\varepsilon)}}}\operatorname{poly}(N)
\]
time for $n\ge2$. Instances with at most one job are solved directly.
The same bound holds for an arbitrary number of identical machines.
\end{proposition}
\begin{proof}
We prove the one-machine bound here; Section~\ref{sec:multi-completion}
establishes the same bound for arbitrary $M$.
Keep $Q=\delta^{-1}$ and define the absolute constant $C_0=22400^2$.
The parameter definitions in~\eqref{eq:constants} give
$\kappa(m)=4Km\delta^{-6}\ge1$ and
$c(\kappa)=1+2\kappa(\kappa+1)\le5\kappa^2$. Hence
\[
 m_{i+1}\le
 \bigl[280Km_i\cdot5(4Km_i\delta^{-6})^2\bigr]^2
 =C_0K^6m_i^6\delta^{-24},
\]
and, since $m_0=1$ and $D(K)=m_2$,
\begin{equation}\label{eq:D-growth}
 D(K)\le C_0^7K^{42}\delta^{-168}.
\end{equation}
The preprocessing cap is
$K_0=(6Q)^{6Q+3}$, so $\log_2K_0=O(Q\log Q)$.
Write $K_\ell=K_0a_\ell$ and $a_1=4$. The least multiple of $a_\ell$
above a real threshold $u\ge a_\ell$ is at most $u+a_\ell\le2u$.
In the scale recurrence the threshold
$4K_\ell D(K_\ell)/\delta^2$ dominates $2a_\ell$. Applying
\eqref{eq:D-growth} therefore yields
\[
 a_{\ell+1}\le
 8C_0^7K_0^{43}Q^{170}a_\ell^{43}
 =B_0a_\ell^{43},
 \qquad \log_2B_0=O(Q\log Q).
\]
Iterating $\log_2a_{\ell+1}\le43\log_2a_\ell+\log_2B_0$ for $Q$
steps gives the explicit estimate
\[
 \log_2a_{Q+1}
 \le 2\cdot43^Q+\frac{43^Q-1}{42}\log_2B_0
 \le43^Q\left(2+\frac{\log_2B_0}{42}\right)
 =2^{O(Q)}.
\]
It follows that all candidate caps $K_\ell$, table orders $D(K_\ell)$,
and the exact-enumeration cutoff
$R=\lceil\max\{(6Q)^3,2K_0a_{Q+1}\}\rceil$ are at most
$2^{2^{O(Q)}}$.

Enumerating all ordered job subsets up to the cutoff costs $n^{O(R)}$
times a polynomial in the binary input length. The local relaxation has
$n^{O(K_\ell D(K_\ell))}$ entries and constraints, and all remaining
computations are polynomial in this explicit representation and $N$.
There are only $Q$ candidates. Parameter construction and factors depending
only on $Q$ can also be absorbed into the displayed bound for $n\ge2$.
Thus the total time is $n^{2^{2^{O(Q)}}}\operatorname{poly}(N)$.
Finally the choice in the theorem proof has
$Q=\lceil100+1000/\bar\varepsilon\rceil=O(1/\bar\varepsilon)$.
\end{proof}

\appendix
\section{Proof of the preprocessing lemma}\label{sec:preprocessing}
We prove Lemma~\ref{lem:preprocessing}. The construction first produces
fine blocks with a constant schedule cap, and then uses nested grouping
scales to accommodate the prescribed function $D$.

\subsection{Greedy refinement}
For an interval $B$ and a supplied set of job identities $E$, write
$\operatorname{Greedy}(B,E)$ for the following subroutine. Make a fresh
working copy of $E$. Repeatedly select from that copy a job with earliest
feasible completion in $B$, starting no earlier than the preceding completion,
and remove the selected identity from this working copy. Stop when no job
in the copy can be appended. The supplied set $E$ is never modified by the
subroutine. All ties are broken deterministically.
The following charging observation will be used
for jobs that remain unselected throughout a run. If $[u,v)$ is a feasible
execution of such a job inside the run's interval, then some greedy
completion belongs to $(u,v]$. Indeed, just after the last greedy completion
at or before $u$, or at the left endpoint if none exists, that execution
is an available candidate. Thus either the next greedy completion is at
most $u$, contradicting its choice as the last one, or it lies in $(u,v]$.
If there is no next selection, the available candidate contradicts
termination. Disjoint executions have disjoint intervals $(u,v]$, so
charging each execution in this way defines an injection into the greedy
completions.

Put $\tau=6Q=6/\delta$, $L_i=\tau^{i+2}$, and $K_0=\tau^{\tau+3}$.
Run $\operatorname{Greedy}([0,T),J)$, and let $J_1$ be its selected
identity set; retain the resulting execution intervals as well.
Cut after every $L_1$ selected jobs, except after the final selected job.
For every iteration $i\ge2$, carry out the following scan.
\begin{enumerate}[label=(\arabic*)]
\item Reset the iteration's availability set to $E_i:=J_{i-1}$ and
initialize $J_i:=\varnothing$. No removal from a previous iteration is
carried over: its selected identities are precisely the new input set.
\item Visit the blocks present at the start of iteration $i$, from left
to right. For a parent block $B$, compute the tentative schedule
$P_B:=\operatorname{Greedy}(B,E_i)$. Its construction modifies only the
subroutine's fresh working copy.
\item If $|P_B|>L_i$, keep its execution intervals, add its identities to
$J_i$, and replace $E_i$ by $E_i\setminus P_B$. Subdivide $B$ after every
$L_i$ selected jobs, except after the final selected job. The resulting
children are not scanned again at this iteration.
\item If $|P_B|\le L_i$, discard the tentative schedule and leave $E_i$
unchanged. In particular, its selected identities can be tried in later
parent blocks of this same iteration.
\end{enumerate}
Here and below, a schedule in a set operation denotes its identity set.
The kept parent schedules are disjoint in time. Their identities are also
disjoint, because a committed identity is removed from $E_i$ before the
next parent run. Their union is therefore a feasible schedule on the
identity set $J_i$. Every identity considered during iteration $i$ belongs
to $J_{i-1}$, proving $J_i\subseteq J_{i-1}$.
Stop at the first iteration with
$|J_{i-1}\setminus J_i|\le|J_1|/\tau$. Such an iteration exists by $i=\tau+1$:
otherwise the $\tau$ disjoint differences for $i=2,\ldots,\tau+1$ would contain
more than $|J_1|$ jobs.

Let $R_1=J\setminus J_1$ and $R_j=J_{j-1}\setminus J_j$ for $j\ge2$.
In any schedule lying within a block after iteration $j$, at most $2L_j$
jobs belong to $R_j$. For $j=1$, all these jobs were unselected, and the
charging observation bounds their count by the at most $L_1$ greedy
completions in the block. For $j\ge2$, each identity in $R_j$ belongs to
the reset set $E_j=J_{j-1}$ and to no schedule committed in iteration $j$.
It is consequently present in $E_j$ at the start of every parent run.
A tentative selection can remove it from that run's working copy, but
discarding the run leaves the iteration-level set $E_j$ unchanged.
In a kept child, all $R_j$ jobs were unselected
in that parent's run, and charge to at most $L_j$ completions in that
child. In an unkept parent, at most $L_j$ jobs are identities of its
tentative schedule; all remaining jobs charge to at most $L_j$ completions.
The bound also holds in all finer blocks produced later.

The number of cuts created throughout the construction is at most
\[
 \sum_{j\ge1}|J_j|/L_j
 \le\mathrm{OPT}\sum_{j\ge1}\tau^{-j-2}
 \le2\mathrm{OPT}/\tau^3.
\]
Take an optimal schedule and remove every execution that crosses a final
cut. At most one execution crosses each cut, so this loses at most
$2\mathrm{OPT}/\tau^3$ jobs. Let $i$ be the stopping iteration. Remove from
this aligned schedule all identities in $J_{i-1}$, clear all kept parents
of iteration $i$, and insert $J_i$ in those parents.
Removing $J_{i-1}$ and inserting $J_i$ has net loss at most
$|J_{i-1}\setminus J_i|\le\mathrm{OPT}/\tau$.
After removing $J_{i-1}$, the remaining identities belong to
$R_1,\ldots,R_{i-1}$. Every kept parent therefore contains at most
\[
 2\sum_{j<i}L_j\le4L_i/\tau
\]
of them. Each kept parent contributes more than $L_i$ jobs to $J_i$;
there are at most $\mathrm{OPT}/L_i$ kept parents. Clearing them loses
at most $4\mathrm{OPT}/\tau$ more jobs. The resulting aligned schedule
$J''$ consequently satisfies
\begin{equation}\label{eq:Jdoubleprime}
 |J''|\ge(1-6/\tau)\mathrm{OPT}=(1-\delta)\mathrm{OPT}.
\end{equation}
Each kept child contains only its inserted jobs, at most $L_i$ of them.
An unkept parent contains at most $4L_i/\tau\le L_i$ remaining jobs.
Thus every final block has at most $L_i\le K_0$ jobs of $J''$.
Denote this fine partition by $\mathcal B_0$ and its size by $b_0$.
If $\mathrm{OPT}\ge \tau^3$, the cut bound gives
\begin{equation}\label{eq:b0}
 b_0\le3\mathrm{OPT}/\tau^3.
\end{equation}
The number of cuts is also $O(n)$, since every $J_j$ has at most $n$
jobs and $\tau$ is a constant.

\subsection{Nested grouping scales}
Recall that $Q=\delta^{-1}$ and set $a_1=4$. Recursively choose
$a_{\ell+1}$ to be the \emph{least} positive multiple of $a_\ell$ satisfying
\[
 a_{\ell+1}\ge\max\{2a_\ell,\ 4K_\ell D(K_\ell)/\delta^2\},
 \qquad K_\ell=K_0a_\ell,
 \quad \ell=1,\ldots,Q.
\]
All these numbers depend only on $\delta$ and the prescribed function $D$.
Let $R=\lceil\max\{\tau^3,2K_0a_{Q+1}\}\rceil$. Enumerate all subsets and
orders through size $R+1$, using earliest-start feasibility tests. If
no feasible schedule of size $R+1$ exists, these tests return an exact
optimum, since any larger schedule would contain such a subset.
Otherwise $\mathrm{OPT}>R$ and the preceding construction gives
\[
 b_0\ge |J''|/K_0\ge(1-\delta)\mathrm{OPT}/K_0\ge a_{Q+1}.
\]

For candidate $\ell$, form blocks from consecutive batches of $a_\ell$
fine blocks and superblocks from batches of $a_{\ell+1}$ fine blocks,
truncating the final batches. Divisibility ensures the partitions are
nested. Every block contains at most $K_\ell$ jobs of $J''$, and the
numbers of blocks and superblocks are at most $2b_0/a_\ell$ and
$2b_0/a_{\ell+1}$, respectively.

Fix an execution of a job in $J''$. If it is ineligible at level $\ell$,
its block is neither boundary block and so lies entirely in its window,
whereas its superblock does not. At every finer level, the superblock
containing this execution is contained in the level-$\ell$ block, hence
inside the window; the execution is eligible. At every coarser level,
the block containing the execution contains its level-$\ell$ superblock.
Because that superblock contains the execution but is not contained in
the window, it contains a release or deadline endpoint in precisely the
half-open endpoint convention defining the boundary blocks. The coarser
block is therefore a boundary block, and the execution is eligible there
as well. Thus an execution violates eligibility at at most one level.

Averaging over the $Q$ candidates gives a level for which deletion of all
ineligible executions loses at most $\delta|J''|$. Its remaining schedule
$J'$ satisfies $|J'|\ge(1-\delta)^2\mathrm{OPT}\ge
(1-2\delta)\mathrm{OPT}$. Furthermore~\eqref{eq:b0} gives
\[
 \frac{b_0}{2\delta}\le\frac{3}{2\delta \tau^3}\mathrm{OPT}
 \le(1-2\delta)\mathrm{OPT}\le |J'|.
\]
Since $a_\ell\ge4$, the chosen level satisfies
\[
 |\mathcal B|\le2b_0/a_\ell\le b_0/2\le\delta|J'|,
 \qquad
 K_\ell D(K_\ell)|\mathcal S|
 \le\frac{2K_\ell D(K_\ell)b_0}{a_{\ell+1}}
 \le\delta^2b_0/2\le\delta^3|J'|.
\]
Output all $Q$ candidates; the schedules $J''$ and $J'$ were used only
for the analysis. The enumeration has constant maximum subset size,
and every greedy pass uses binary arithmetic on input endpoints and
processing times. This proves both the running-time bound and all
claims of Lemma~\ref{lem:preprocessing}.

\section{Separation of two boundary graphs}\label{sec:separator}
We use the planar separator theorem of Lipton and Tarjan~\cite{LT79} in its unweighted form: an $n$-vertex planar graph has a vertex separator of size at most $\sqrt{8n}$ whose deletion leaves components of size at most $2n/3$. Such a separator is computable in polynomial time. We first turn this balanced separator into a bound on all remaining components. We then transfer that bound from a planarization back to the block vertices, while deleting complete old groups. The constants below leave slack in the later loss budget.

\begin{lemma}[Recursive planar separation]\label{lem:planar-recursive}
Let $P$ be a planar graph on $p$ vertices and let $M\geq1$ be an integer. In polynomial time one can find a set $Z\subseteq V(P)$ with
\[
 |Z|\leq \frac{35p}{\sqrt M}
\]
such that every component of $P-Z$ has at most $M$ vertices.
\end{lemma}
\begin{proof}
First consider a connected planar graph on $n$ vertices. Apply the planar separator theorem once. If a remaining component has more than $n/2$ vertices, it is unique; its size is at most $2n/3$. Apply the theorem again to that component. Every component resulting from the second application has at most $4n/9$ vertices. Thus the union of the two separators is a half-balanced separator, and its size is at most
\[
 \sqrt{8n}+\sqrt{8(2n/3)}<10\sqrt n.
\]
If no second application is needed, the same bound holds.

Apply this half-balanced separation recursively to every component with more than $M$ vertices. Charge the deletion of at most $10\sqrt u$ vertices from a processed $u$-vertex component evenly to its $u$ vertices, at rate $10/\sqrt u$. For a fixed original vertex, the processed components containing it form a chain; the size of each child is at most half its parent's size. Reading the chain backward from its last component, whose size exceeds $M$, we bound the total charge to the fixed vertex by
\[
 \frac{10}{\sqrt M}\sum_{i=0}^{\infty}2^{-i/2}
 =\frac{10}{(1-2^{-1/2})\sqrt M}
 <\frac{35}{\sqrt M}.
\]
Summing over the original vertices bounds the total number deleted. The recursion has at most $p$ nonempty leaves and polynomially many calls. Applying the construction independently to the initial components handles disconnected graphs.
\end{proof}

Two edges with four distinct endpoints in a cyclic order \emph{cross} if their endpoints strictly alternate. Edges sharing an endpoint do not cross. For an integer $\kappa\geq0$, put
\[
 c(\kappa)=1+2\kappa(\kappa+1).
\]
\begin{lemma}[Two-page separation]\label{lem:two-page}
Let $b\geq0$, $m\geq1$, and $\kappa\geq0$ be integers, and let $G_0,G_1$ be simple graphs on the same $b$ cyclically ordered vertices. Suppose that, within each graph, every edge crosses at most $\kappa$ other edges. Suppose also that every connected component of $G_0$ has at most $m$ vertices. For every integer $M\geq1$, one can delete a union of whole $G_0$-components, containing at most
\[
 \frac{140c(\kappa)mb}{\sqrt M}
\]
vertices, so that every remaining component of $G_0\cup G_1$ has at most $M$ vertices. The construction takes polynomial time.
\end{lemma}
\begin{proof}
\textbf{A planar representation of the two families.}
Draw $G_0$ and $G_1$ on opposite hemispheres of a sphere, placing their common vertices on the equator in the prescribed cyclic order. On each hemisphere use a chord drawing; a sufficiently small perturbation makes all crossings pairwise, without changing which pairs of edges cross. Intersections between edges from different graphs occur only at their common endpoints on the equator.

For either graph, form the graph whose vertices are its edges and whose adjacencies are crossings. Its maximum degree is at most $\kappa$, so a greedy coloring uses at most $\kappa+1$ colors. Edges of one color give a simple outerplanar graph on the $b$ original vertices, with at most $2b$ edges. Consequently each hemisphere has at most $2(\kappa+1)b$ edges and at most $\kappa(\kappa+1)b$ crossings. The planarization $P$, obtained by introducing one vertex per crossing, has at most $c(\kappa)b$ vertices. Parallel edges in $P$ may be suppressed, since they do not change its vertex separators or its connected components.

\smallskip\noindent\textbf{Deleting whole old components.}
Apply Lemma~\ref{lem:planar-recursive} to $P$. Replace every deleted crossing vertex by the endpoints of the two original edges passing through that crossing; retain every deleted original vertex. The resulting set of original vertices has size at most four times the planar separator's size. Delete all $G_0$-components meeting this set. Since each such component has at most $m$ vertices, the number of original vertices deleted is at most $140c(\kappa)mb/\sqrt M$.

\smallskip\noindent\textbf{Why the surviving union is small.}
For every surviving original edge, its subdivided path in $P$ avoids the planar separator: otherwise either an endpoint or all endpoints of an edge through a deleted crossing would have been removed. Every connected component of the surviving union therefore maps into one component of the separated planarization. Such a component contains at most $M$ planar vertices, and hence at most $M$ original vertices. This proves the required bound. The drawings, crossing enumeration, planarization, separator recursion, and component deletions all have polynomial size and can be constructed in polynomial time.
\end{proof}

\section{An arbitrary number of identical machines}
\label{sec:multiple-machines}
Let $M$ be the number of machines and $\mathrm{OPT}_M$ the optimum.
We use a common time partition for all machines. For bounded $M$, each
block configuration records a feasible schedule on all $M$ machines,
and each sampled slot retains its machine label. For sufficiently large
$M$, the algorithm of Im, Li, and Moseley~\cite{ILM20} gives the required
guarantee. This bounded/large-machine division is also used by
Armbruster et al.~\cite[Section~4]{AGTW26}.

\subsection{Preprocessing for a bounded number of machines}
\label{sec:multi-preprocessing}
Fix a positive integer $M$. We extend Lemma~\ref{lem:preprocessing} to
$M$ identical machines, with a cap $K$ on the total number of jobs in a
block across all machines. Its constants may depend on $M$. The
block--superblock partition is common to all machines. Write
$\mathrm{OPT}_M$ for the optimum on $M$ machines.

\begin{lemma}\label{lem:multi-preprocessing}
For every fixed positive integer $M$ and every prescribed computable
integer-valued function $D(K)\ge1$ depending only on $K$ and $\delta$,
a deterministic algorithm either solves the $M$-machine instance exactly
or constructs $Q=\delta^{-1}$ candidate block--superblock partitions,
with caps depending only on $\delta$, $M$, and $D$. At least one candidate
admits an eligible $M$-machine schedule $J'$ with at most its cap $K$ jobs
per block and
\[
 |J'|\ge(1-2\delta)\mathrm{OPT}_M,\qquad
 |\mathcal B|\le\delta|J'|,\qquad
 KD(K)|\mathcal S|\le\delta^3|J'|.
\]
The algorithm runs in $n^{O_{\delta,M}(1)}\operatorname{poly}(N)$ time,
where the hidden constant may depend on $D$.
\end{lemma}
\begin{proof}
Run the greedy refinement of Appendix~\ref{sec:preprocessing} with
\[
 \tau=6MQ,\qquad L_i=\tau^{i+2},\qquad K_0=\tau^{\tau+3}.
\]
Each invocation of $\operatorname{Greedy}(B,E)$ still constructs a
\emph{one-machine} schedule, with the same fresh working copy of $E$ and
the same iteration-level reset and commitment rules. Thus the sets $J_i$
remain nested, each $J_i$ has a feasible one-machine schedule, and the
stopping iteration is at most $\tau+1$.

The charging observation in that appendix now has multiplicity at most
$M$. Indeed, restrict an $M$-machine schedule of identities unselected by
a greedy run to one machine. Its execution intervals are disjoint, so
their charges to greedy completions are injective. Summing over the $M$
machines gives the claim. Consequently, inside any block after iteration
$j$, at most $2ML_j$ scheduled identities belong to $R_j$. In a kept child,
all these identities were unselected and charge to at most $L_j$
completions with multiplicity at most $M$. In an unkept parent, at most
$L_j$ identities belong to its tentative greedy schedule; all other
identities charge to its at most $L_j$ completions. This gives
$(M+1)L_j\le 2ML_j$. The same bounds persist in finer blocks.

The number of cuts is at most
\[
 \sum_{j\ge1}\frac{|J_j|}{L_j}
 \le \frac{2\mathrm{OPT}_M}{\tau^3}.
\]
Deleting every optimal execution that crosses a cut loses at most
$2M\mathrm{OPT}_M/\tau^3$ jobs, because at most one execution on each
machine crosses a cut. Let $i$ be the stopping iteration. Remove all
identities in $J_{i-1}$, clear all machines inside every kept parent of
iteration $i$, and insert its one-machine greedy schedule on the first
machine. The deletion of $J_{i-1}$ and insertion of $J_i$ have net loss at
most $\mathrm{OPT}_M/\tau$. After deleting $J_{i-1}$, each kept parent
contains at most
\[
 2M\sum_{j<i}L_j\le\frac{4ML_i}{\tau}
\]
remaining jobs. There are at most $\mathrm{OPT}_M/L_i$ kept parents, so
clearing them loses at most $4M\mathrm{OPT}_M/\tau$ jobs. The resulting
aligned schedule $J''$ satisfies
\[
 |J''|\ge
 \left(1-\frac{2M}{\tau^3}-\frac{1+4M}{\tau}\right)\mathrm{OPT}_M
 \ge\left(1-\frac{6M}{\tau}\right)\mathrm{OPT}_M
 =(1-\delta)\mathrm{OPT}_M.
\]
Every kept child contains at most $L_i$ inserted jobs, and every unkept
parent contains at most $4ML_i/\tau\le L_i$ remaining jobs. Thus the cap
is $K_0$. If $\mathrm{OPT}_M\ge\tau^3$, the number $b_0$ of fine blocks
satisfies $b_0\le3\mathrm{OPT}_M/\tau^3$. The construction has $O(n)$
blocks and uses binary arithmetic on the original endpoints.

Apply the nested grouping scales from Appendix~\ref{sec:preprocessing}
with this value of $K_0$ and the same $Q=\delta^{-1}$. To handle the
small-optimum branch, let
$R=\lceil\max\{\tau^3,2K_0a_{Q+1}\}\rceil$ and enumerate job subsets
of size at most $R+1$, assignments to machines, and orders on each
machine. For each such choice, earliest-start schedules test feasibility
using binary arithmetic. If no feasible subset of size $R+1$ exists, this
enumeration finds an optimum. Otherwise the lower bound on $b_0$ and the
nested-scale eligibility argument apply unchanged: eligibility depends
on the job's window and execution interval. In particular, each execution
is ineligible at at most one of the $Q$ levels. A good level has an
eligible $M$-machine schedule $J'$ with
\[
 |J'|\ge(1-2\delta)\mathrm{OPT}_M,\qquad
 |\mathcal B|\le\delta|J'|,\qquad
 KD(K)|\mathcal S|\le\delta^3|J'|.
\]
Here $K$ bounds the total number of jobs per block. The last two
inequalities follow from the same calculation as in the single-machine
case, since $\tau\ge6Q$. For fixed $M$ the running time is
$n^{O_{\delta,M}(1)}\operatorname{poly}(N)$.
\end{proof}

\subsection{Configurations and rounding on a fixed number of machines}
Fix a constant number $M$ of identical machines and apply the preprocessing
construction above. The resulting blocks form a common partition of the
time horizon for all machines. The cap $K$ bounds the total number of jobs
in a block, summed over the $M$ machines.

For each block $B$, let $\mathcal C_B$ contain the eligible job sets of
size at most $K$ that admit a schedule on the $M$ machines inside $B$.
Store one such schedule for each set, including the machine assigned to
each job. These configurations can be enumerated by assigning each
candidate job to one of the $M$ machines and enumerating the order on
each machine. For a fixed assignment and orders, schedule each machine's
jobs successively at their earliest feasible start times in $B$.
The test succeeds for some assignment and orders exactly when the set is
feasible. Since $M$ and $K$ are constants, this takes
$n^{O(K)}\operatorname{poly}(N)$ time per block.

The distributions $x_{B,\cdot}$ and $\mu_U$, their consistency constraints,
and the job constraints are the same as in \cref{sec:relaxation}.
Every tuple in $\mathcal T_U$ contains each job identity at most once
across all its blocks and machines. Restricting the eligible schedule
provided by preprocessing to each table gives a feasible integral point.
Consequently the candidate-selection rule still yields
\eqref{eq:LPbounds}, with $\mathrm{OPT}=\mathrm{OPT}_M$.

The boundary blocks of a job depend only on its release time and deadline.
In particular, each nonlocal job still has precisely two boundary blocks.
Within a sampled block or group its identity occurs at most
once. Thus the quantities $a_j,b_j,c_j,h_j,q_j$ and the reservations in
\eqref{eq:reservation} have the same interpretation. A group of at most
$m$ blocks contains at most $\beta=Km$ job occurrences. The bounds on
active jobs incident to a group, the protected crossing degree, and the
mass lost when blocks are deleted therefore remain unchanged. This proves
the two coordination inequalities in \eqref{eq:coordination}.

Give every sampled slot its stored machine label. On each machine the
slots are pairwise disjoint: this holds within a configuration, and
different blocks are disjoint time intervals. A compatible job assigned
to a slot $[s,t)$ runs on that slot's machine from
$\max\{s,r_j\}$ for $p_j$ time units. The compatibility condition
\eqref{eq:compatibility} places its execution inside the slot and inside
its window. A matching consequently produces a feasible schedule on the
$M$ machines.

The fractional-matching analysis also applies to these labelled slots.
The flags for boundary absence and duplication are determined by the
same occurrence records in at most two groups, so the variance bound
uses the same $\beta$. For interior assignments, every recipient window
contains the supplying superblock and its processing time fits the
supplying slot. For a boundary repair, the crossing relation places the
selected donor boundary slot inside the recipient window and $p_j\le p_k$ ensures that
the recipient fits. Identical machines allow either assignment on the
machine of the chosen slot. The capacity reservations and the donor's
remaining baseline assignment concern individual jobs and slots, so all
capacity calculations in \cref{sec:matching} are unchanged. Hence the
expected maximum matching cardinality is at least
$F'-W+H-4\delta F$.

Combining the three scores as in \cref{sec:main-proof}, and using the same
finite-bit implementation, gives an expected cardinality of at least
$(11/12-8\delta)\mathrm{OPT}_M$. With the choice of $\delta$ made there,
this is at least $\mathrm{OPT}_M/(12/11+\varepsilon)$.

\subsection{An arbitrary number of machines}
\label{sec:multi-completion}
Im, Li, and Moseley~\cite{ILM20} give, for every fixed $\eta>0$, a
polynomial-time algorithm for unweighted throughput maximization on $M$
identical machines whose expected retained fraction is at least
\[
  1-C\sqrt{\frac{\log M}{M}}-\eta,
\]
for an absolute constant $C\ge1$ and $M\ge2$, where $\log$ denotes the
natural logarithm. Their unweighted result
allows binary-encoded release times, deadlines, and processing times.
We use this result for a sufficiently large number of machines, as in
the extension of Armbruster et al.~\cite[Section~4]{AGTW26}.

Fix $\eta=1/32$ and the absolute integer
\[
 M_\star=\max\bigl\{2,\lceil(32C)^4\rceil\bigr\}.
\]
For $M\ge M_\star$, the inequality $\log M\le\sqrt M$ gives
\[
 C\sqrt{\frac{\log M}{M}}\le C M^{-1/4}\le\frac1{32}.
\]
The algorithm of Im, Li, and Moseley therefore returns a feasible schedule
with expected cardinality at least
\[
 \frac{15}{16}\mathrm{OPT}_M
 \ge\frac{11}{12}\mathrm{OPT}_M
 \ge\frac{\mathrm{OPT}_M}{12/11+\varepsilon}.
\]
For $M<M_\star$, apply the fixed-machine construction in the preceding
subsections. This proves the approximation guarantee in
Theorem~\ref{thm:main} for every number of machines.

We may replace an input machine count larger than $n$ by $n$, since any
schedule of at most $n$ jobs uses at most $n$ machines. Instances with at
most one job are solved directly. For $M<M_\star$, the preprocessing
parameters satisfy
\[
 \tau=6MQ=O(Q),\qquad
 K_0=(6MQ)^{6MQ+3},\qquad \log_2 K_0=O(Q\log Q),
\]
uniformly over this range of $M$. The bound
$D(K)\le C_0^7K^{42}\delta^{-168}$ and the recurrence for the nested
scales are unchanged. The proof of Proposition~\ref{prop:runtime}
therefore gives the same bound
\[
 n^{\,2^{\,2^{O(1/\bar\varepsilon)}}}\operatorname{poly}(N),
 \qquad \bar\varepsilon=\min\{\varepsilon,1\}.
\]
The additional factor $M^K$ for machine assignments is absorbed into
$n^{O(K)}$ because $M<M_\star$ and $n\ge2$. For $M\ge M_\star$, the algorithm invoked
at the fixed accuracy $\eta=1/32$ runs in polynomial time in the binary
input length. The displayed bound consequently holds for the complete
algorithm with an arbitrary input machine count.

\section*{Acknowledgment}
ChatGPT 6 Pro was used to assist in developing the mathematical proofs and preparing the manuscript.

\end{document}